\documentclass[12pt]{article}
\usepackage{a4wide}
\usepackage{latexsym}
\usepackage{amsmath}
\usepackage{amsfonts}
\usepackage{amscd}
\usepackage{cite}
\usepackage{graphicx}
\usepackage{axodraw}
\usepackage{float}

\usepackage{pslatex}
\usepackage[latin1]{inputenc}
\usepackage[OT2,T1]{fontenc}

\newcommand{\bq}{\begin{eqnarray}}
\newcommand{\eq}{\end{eqnarray}}

\DeclareSymbolFont{cyrletters}{OT2}{wncyr}{m}{n}
\DeclareMathSymbol{\Sha}{\mathalpha}{cyrletters}{"58}

\begin{document}

\thispagestyle{empty}

\begin{flushright}
  MITP/16-142
\end{flushright}

\vspace{1.5cm}

\begin{center}
  {\Large\bf The matrix element method at next-to-leading order for arbitrary jet algorithms\\
  }
  \vspace{1cm}
  {\large Robin Baumeister and Stefan Weinzierl\\
\vspace{2mm}
      {\small \em PRISMA Cluster of Excellence, Institut f{\"u}r Physik, }\\
      {\small \em Johannes Gutenberg-Universit{\"a}t Mainz,}\\
      {\small \em D - 55099 Mainz, Germany}\\
  } 
\end{center}

\vspace{2cm}

\begin{abstract}\noindent
  {
The matrix element method usually employs leading-order matrix elements.
We discuss the generalisation towards higher orders in perturbation theory
and show how the matrix element method can be used at next-to-leading order 
for arbitrary infrared-safe jet algorithms.
We discuss three variants at next-to-leading order.
The first two variants work at the level of the jet momenta.
The first variant adheres to strict fixed-order in perturbation theory.
We present a method for the required integration over the radiation phase space.
The second variant is inspired by the POWHEG method and works as the first variant
at the level of the jet momenta.
The third variant is a more exclusive POWHEG version.
Here we resolve exactly one jet into two sub-jets.
If the two sub-jets are resolved above a scale $p_\bot^{\mathrm{min}}$, the likelihood
is computed from the POWHEG-modified real emission part, otherwise it is given
by the POWHEG-modified virtual part.
   }
\end{abstract}

\vspace*{\fill}

\newpage

\section{Introduction}
\label{sect:intro}

Precision particle physics relies on our ability to extract the fundamental parameters of the theory (like couplings and masses)
from the experimental data.
The matrix element method \cite{Kondo:1988yd,Kondo:1991dw,Kondo:1993in,Dalitz:1991wa,Artoisenet:2010cn,Brandt:2014tta}
is a very helpful tool to this aim.
It allows us to make maximal use of all available kinematic information for each individual event.
A prominent application of the matrix element method is the determination of the top mass.
For a review of the experimental aspects of the matrix element method with an emphasis on 
top mass measurements we refer to \cite{Fiedler:2010sg}.

Usually these analyses are based on leading-order matrix elements.
With increasing precision one would like to extend the matrix element method to higher orders in perturbation theory.
First steps in this direction were done in \cite{Alwall:2010cq,Campbell:2012cz,Campbell:2013hz,Soper:2014rya,Martini:2015fsa}.
In particular, in \cite{Campbell:2012cz,Campbell:2013hz} the extension of the matrix element method to 
next-to-leading order (NLO) for processes with colourless final states was presented.
Coloured final states were considered in \cite{Martini:2015fsa}, however the technique presented there 
is based on a very special $3 \rightarrow 2$ jet clustering algorithm.
It is desirable to extend the matrix element method at NLO to arbitrary infrared-safe jet algorithms.

In this paper we solve this problem and describe 
how the matrix element method can be used at NLO for general processes and general (infrared-safe) observables.
The typical application is a process with hadronic final states, where jets are defined by an arbitrary infrared-safe jet algorithm.
The evaluation of the likelihood at NLO requires the integration over the radiation phase space of the real emission.
We show how this integration can be done numerically for an arbitrary infrared-safe jet algorithm.
We present three alternative variants of the matrix element method at NLO. 
The differences among these three variants are on the one hand 
related to how smearing effects due to imperfect detector resolutions are implemented in the transfer function
and on the other hand related to the used matrix elements, either strictly next-to-leading order or POWHEG-modified matrix elements.

Within the first variant, which we may call ``strict fixed-order'', we first cluster in the theoretical perturbative calculation
the partons into jets. Smearing effects are then applied to the jet momenta.
The matrix elements are -- as the name implies -- the strict NLO matrix elements.

The second variant is a small modification of the first variant.
The second variant works as the first variant on the level of the jets.
However, we take now the matrix elements to be the POWHEG-modified matrix elements, i.e. matrix elements where Sudakov factors are included.
For a perfect detector and an infrared-safe observable the first two variants agree at NLO, numerical differences are due to higher-order
effects (entering through the Sudakov factor within the POWHEG method).

The third variant is a more exclusive POWHEG version.
Here we resolve exactly one jet into two sub-jets.
If the two sub-jets are resolved above a scale $p_\bot^{\mathrm{min}}$, the likelihood
is computed from the POWHEG-modified real emission part, otherwise it is given
by the POWHEG-modified virtual part.

This paper is organised as follows:
In the next section we give an overview of the matrix element method.
Section~\ref{sect:transfer_function} is devoted to the transfer function.
In section~\ref{sect:nlo} we describe the essential ingredients of a NLO calculation.
Section~\ref{sect:MEM_NLO} contains the main results of this article: We present the three variants
(``strict fixed-order'', ``POWHEG-inspired'' and ``sub-jet based'') of the matrix element method at NLO.
Finally, section~\ref{sect:conclusions} contains our conclusions.

\section{The matrix element method}
\label{sect:matrix_element_method}

Let us consider a theory, depending on parameters which we assemble in a vector $\vec{\alpha}$.
The entries of $\vec{\alpha}$ are called model parameters.
Typical examples are couplings or masses.
Let us denote by $\vec{x}$ the experimentally measured variables in an event.
Typically, $\vec{x}$ consists of jet energies, jet rapidities and the azimuthal angles of the jets.
We further denote by $\vec{y}$ the corresponding partonic variables of a single event within a perturbative calculation.
Within a leading-order calculation a jet is modelled by one parton and the jet momenta coincide with the parton momenta.
Thus in typical applications at leading-order we may take the variables $\vec{y}$ as the theoretical jet momenta (or equivalently as the parton
momenta).
Identifying a jet with a single parton is a very crude approximation.
This approximation is improved by including higher-orders from perturbation theory.
Starting from NLO, a jet may be modelled by more than one parton.
This implies that at NLO we have to distinguish between jet momenta and parton momenta.

Let $r$ and $s$ be the dimensions of the vectors $\vec{x}$ and $\vec{y}$, respectively.
Note that the dimensions of $\vec{x}$ and $\vec{y}$ need not to be the same.
We denote by
\bq
 \frac{d^s\sigma}{dy_1 ... dy_s}
\eq
the differential cross section. 
The total cross section is given by
\bq
 \sigma
 & = &
 \int d^sy
 \frac{d^s\sigma}{dy_1 ... dy_s}.
\eq
At leading-order the differential cross section is given by
\bq
\label{leading_order_cross_section}
 d\sigma^{\mathrm{LO}}
 & = &
 \sum\limits_{f_a,f_b}
 \int dx_a \int dx_b  
             \frac{f_{f_a}(x_a) f_{f_b}(x_b)}{2 \hat{s} n_s(a) n_s(b) n_c(a) n_c(b)}
             \sum\limits_{\mathrm{spins,colour}} 
             \left| {\mathcal A}^{(0)} \right|^2
             d\phi_{n}.
\eq
The sum involving $f_a$ and $f_b$ is over the flavours of the two initial state partons,
$x_a$ and $x_b$ denote as usual the momentum fractions of the two initial state partons.
The variable $\hat{s}$ denotes the partonic centre-of-mass energy, $n_s(i)$ and $n_c(i)$ give 
for parton $i$ the number
of spin degrees of freedom and the number of colour degrees of freedom, respectively.
The relevant Born matrix element is denoted by ${\mathcal A}^{(0)}$.
The quantity $d\phi_n$ stands for the phase space measure of $n$ external particles.
The phase space for $n$ final state particles is $(3n-4)$-dimensional.
We denote the $n$-particle phase space by $\Phi_n$ and a point in the phase space by $\phi_n$.
It will be convenient to abbreviate eq.~(\ref{leading_order_cross_section}) as
\bq
\label{short_notation_LO}
 d\sigma^{\mathrm{LO}}
 & = &
 B\left(\phi_n\right) d\phi_n.
\eq
The model-dependent likelihood to observe an event $\vec{x}$ for model parameters $\vec{\alpha}$ is given by
\bq
\label{def_likelihood}
 {\mathcal L}\left(\vec{x}|\vec{\alpha}\right)
 & = &
 \frac{1}{\sigma} \int d^sy \frac{d^s\sigma}{dy_1 ... dy_s} W\left(\vec{x},\vec{y}\right).
\eq
The matrix elements enter in the partonic differential cross section $d\sigma$.
At leading order, the differential cross section $d\sigma^{\mathrm{LO}}/(dy_1 ... dy_s)$ is a non-negative function,
as required for the interpretation of the left-hand side of eq.~(\ref{def_likelihood}) 
as a likelihood function.
If eq.~(\ref{def_likelihood}) is used with leading-order matrix elements, we assumed implicitly that the leading-order
matrix elements are integrable over the $n$-particle phase space.
If this is not the case, one adds a jet function. This is discussed in more detail in section~(\ref{sect:nlo}).

The function $W(\vec{x},\vec{y})$ is called the transfer function and gives the probability that 
a partonic event $\vec{y}$ is measured in the detector as an event $\vec{x}$.
The transfer function satisfies
\bq
 \int d^rx \; W\left(\vec{x},\vec{y}\right)
 & = & 1.
\eq
For a set of events $\{\vec{x}\} = \{\vec{x}_1, ..., \vec{x}_N\}$ one defines the likelihood function for this set
as
\bq
\label{likelihood_for_set_of_events}
 {\mathcal L}\left(\left\{\vec{x}\right\}|\vec{\alpha}\right)
 & = &
 \prod\limits_{i=1}^N
 {\mathcal L}\left(\vec{x}_i|\vec{\alpha}\right).
\eq
The best fit for the model parameters $\vec{\alpha}$ is given by the values, which
maximise ${\mathcal L}(\{\vec{x}\}|\vec{\alpha})$, or equivalently maximise
\bq
 \ln {\mathcal L}\left(\left\{\vec{x}\right\}|\vec{\alpha}\right).
\eq
Contours of $a$ standard deviations for the model parameters $\vec{\alpha}$
are obtained from the equation
\bq
 \ln {\mathcal L}\left(\left\{\vec{x}\right\}|\vec{\alpha}\right)
 & = & 
 \ln {\mathcal L}\left(\left\{\vec{x}\right\}|\vec{\alpha}_{\mathrm{max}}\right)
 - \frac{a^2}{2}.
\eq

\section{The transfer function}
\label{sect:transfer_function}

Let us briefly recapitulate our theoretical understanding of high-energy scattering events
in had\-ron-hadron collisions.
We start from the hard scattering event. The hard scattering event may be calculated reliably in perturbation
theory and involves only a few partons.
The hard scattering is followed by parton showering, hadronisation and the decay of unstable particles.
In addition to the particles originating from the hard scattering there will be particles originating
from the interactions of the hadron remnants.
Soft particles from the hadron remnants constitute the underlying event,
hard particles from the hadron remnants are referred to as multiple interactions.
Furthermore there might be more than one hadron-hadron scattering within a bunch crossing.
This is known as pile-up events.
The physical particles (photons, electrons, muons, mesons, hadrons, etc.) are then
detected in the detector.
Typical detectors are multi-purpose detectors, consisting of sub-systems for tracking, electro-magnetic and
hadronic calorimeters and muon chambers.
The raw data from the detector is then passed through the trigger, followed by a reconstruction
of the jet momenta.

The transfer function $W(\vec{x},\vec{y})$ describes the conditional probability to observe a
detector-level event $\vec{x}$, given a certain partonic event $\vec{y}$.
The transfer function models detector effects.
As already mentioned, the dimensions of the vectors $\vec{x}$ and $\vec{y}$ need not to be the same.
For example, this will be the case if one or more particles escape detection (like neutrinos), such that there is not enough information
to reconstruct all kinematic variables.
The momenta of all particles (detected and undetected) will appear in the partonic variables
$\vec{y}$, however the variables corresponding to the missing information will be absent in the detector-level variables $\vec{x}$.
In the following we will assume that $s \ge r$ and we recall that
$s = \dim \vec{y}$ and $r = \dim \vec{x}$.
It is beneficial to choose the two sets of variables $\vec{x}$ and $\vec{y}$ as closely related as possible.
By an appropriate choice of the two sets of variables $\vec{x}$ and $\vec{y}$ it is often possible to assume
that the transfer function factorises 
\bq
 W\left(\vec{x},\vec{y}\right)
 & = &
 \prod\limits_{i=1}^r W_i\left(x_i,y_i\right).
\eq
This assumption implies that the $y$-variables $y_{r+1}$, ..., $y_s$, which have no partner in the
$x$-variables, are marginalised.
For the modelling of the individual functions $W_i(x_i,y_i)$ one chooses in practical applications
often a simple delta distribution
\bq
 W_i\left(x_i,y_i\right)
 & = & 
 \delta\left(x_i-y_i\right),
\eq
or a Gaussian distribution
\bq
 W_i\left(x_i,y_i\right)
 & = & 
 \frac{1}{\sigma_i \sqrt{2\pi}}
 e^{-\frac{1}{2} \left(\frac{x_i-y_i}{\sigma_i}\right)^2}.
\eq
The delta distribution may be viewed as the limit $\sigma \rightarrow 0+$ of a Gaussian distribution:
\bq
 \delta\left( x-y \right)
 & = &
 \lim\limits_{\sigma \rightarrow 0+}
 \frac{1}{\sigma \sqrt{2\pi}}
 e^{-\frac{1}{2} \left(\frac{x-y}{\sigma}\right)^2}
\eq
Of course, more sophisticated models for the transfer function are possible, 
but the cases discussed above suffice for our purpose.

\section{Next-to-leading order calculations}
\label{sect:nlo}

In the notation of eq.~(\ref{short_notation_LO}) we may write the differential NLO cross section as
\bq
\label{short_notation_NLO}
 d\sigma^{\mathrm{NLO}}
 & = &
 \left[ B\left(\phi_n\right) + V\left(\phi_n\right) + C\left(\phi_n\right) \right] d\phi_n
 +
 R\left(\phi_{n+1}\right) d\phi_{n+1},
\eq
where $V(\phi_n)$ denotes the renormalised virtual contribution, $R(\phi_{n+1})$ the real contribution
and $C(\phi_n)$ a counter term for initial state collinear singularities.
The terms in the square bracket live on the phase space of $n$ final state particles,
while the real emission contribution lives on the phase space of $(n+1)$ final state particles.
The two contributions are individually divergent, only their sum is finite.
In order to render the individual contributions finite, one either employs
phase space slicing \cite{Giele:1992vf,Giele:1993dj,Keller:1998tf,Harris:2001sx,Gao:2012ja,Boughezal:2015dva,Gaunt:2015pea}
or 
the subtraction method \cite{Kunszt:1994mc,Frixione:1996ms,Catani:1997vz,Dittmaier:1999mb,Phaf:2001gc,Catani:2002hc,Weinzierl:2005dd,Frederix:2009yq,Frixione:2011kh,Kosower:1998zr,Campbell:1998nn,Gehrmann-DeRidder:2005cm,Daleo:2006xa,Somogyi:2006cz,Nagy:2007ty,Chung:2010fx,Dittmaier:2008md,Czakon:2009ss,Gotz:2012zz,Bevilacqua:2013iha}.
Within the subtraction method
one subtracts and adds a suitable approximation term $A(\phi_{n+1})$ and rewrites eq.~(\ref{short_notation_NLO})
as
\bq
\label{subtracted_NLO}
 d\sigma^{\mathrm{NLO}}
 = 
 \left[ B\left(\phi_n\right) + V\left(\phi_n\right) + C\left(\phi_n\right) + A\left(\phi_{n+1}\right) d\phi_{\mathrm{unres}} \right] d\phi_n
 +
 \left[ R\left(\phi_{n+1}\right) - A\left(\phi_{n+1}\right) \right] d\phi_{n+1}.
\eq
Here we used the fact that we may write the $(n+1)$-particle phase space as a product of a $n$-particle phase space
and a radiation phase space (also called unresolved phase space):
\bq
 d\phi_{n+1}
 & = &
 d\phi_n d\phi_{\mathrm{unres}}.
\eq
Within the subtraction method one defines in addition for each subtraction term a projection from the $(n+1)$-particle phase
space to the $n$-particle phase space, which we denote by
\bq 
\label{projection_to_Born}
 \phi_n & = & \pi^{(\alpha)}\left(\phi_{n+1}\right),
\eq
where $\alpha$ labels the individual subtraction terms.

For our purpose it will be more convenient to use phase space slicing.
The phase space slicing approach has recently seen a 
revival in the form of $n$-jettiness slicing \cite{Stewart:2010tn,Gao:2012ja,Boughezal:2015dva,Gaunt:2015pea}.
One introduces a small parameter $\tau_n^{\mathrm{min}}$ and divides the $(n+1)$ phase space into the two regions
$\tau_n > \tau_n^{\mathrm{min}}$ and $\tau_n < \tau_n^{\mathrm{min}}$.
The former region is free of singularities and can be integrated numerically.
The latter region contains all infrared singularities. Here, one approximates the real emission matrix element with
its soft and collinear limits. This introduces an error of order ${\mathcal O}(\tau_n^{\mathrm{min}})$.
By choosing $\tau_n^{\mathrm{min}}$ small enough one ensures that this approximation error can be neglected.
Within the slicing method we have
\bq
\label{slicing_NLO}
 d\sigma^{\mathrm{NLO}}
 & = &
 \left[ B\left(\phi_n\right) + V\left(\phi_n\right) + C\left(\phi_n\right) + I\left(\phi_n\right) \right] d\phi_n
 +
 \theta\left(\tau_n-\tau_n^{\mathrm{min}}\right)
 R\left(\phi_{n+1}\right) d\phi_{n+1},
\eq
where $I\left(\phi_n\right)$ denotes the integral of the soft and collinear approximation term over the unresolved region 
$\tau_n<\tau_n^{\mathrm{min}}$.
The theta function ensures that the real contribution is restricted to the resolved region 
$\tau_n>\tau_n^{\mathrm{min}}$.

Let us now consider an infrared-safe jet algorithm.
Typical examples used in hadron collisions are the $k_\perp$-algorithm \cite{Stirling:1991ds,Catani:1993hr,Ellis:1993tq},
the SISCone algorithm \cite{Salam:2007xv} or the anti-$k_\perp$-algorithm \cite{Cacciari:2008gp}.
A jet algorithm defines a jet function $\Theta_n$, which equals one if the (partonic) event is classified as an $n$-jet event
and zero otherwise.
The $n$-jet cross section is given within the phase space slicing approach by
\bq
\label{nlo_cross_section}
 \sigma_n^{\mathrm{NLO}}
 & = &
 \int \Theta_n\left(\phi_n\right)
 \left[ B\left(\phi_n\right) + V\left(\phi_n\right) + C\left(\phi_n\right) + I\left(\phi_n\right) \right] d\phi_n
 \nonumber \\
 & &
 +
 \int
 \Theta_n\left(\phi_{n+1}\right) \theta\left(\tau_n-\tau_n^{\mathrm{min}}\right) R\left(\phi_{n+1}\right) 
 d\phi_{n+1}.
\eq
Note that the jet function is evaluated either with $n$-particle kinematics $\phi_n$ 
or $(n+1)$-particle kinematics $\phi_{n+1}$.
The latter situation occurs in the real emission term.

In addition, a jet algorithm clusters the parton momenta to jet momenta.
We assume that each jet is characterised by three variables, typically the jet energy and two angles, subject
to the constraints imposed by momentum conservation:
For hadron-hadron collisions we have momentum conservation in the transverse plane, giving us two constraints.
In electron-positron annihilation all four components of the sum of the four-momenta of the two incoming
particles are conserved, giving us four constraints.
We assemble the variables describing a $n$-jet configuration in a vector $\vec{j}$.
The dimension of the vector $\vec{j}$ is $(3n-2)$ for hadron-hadron collisions and $(3n-4)$ for 
electron-positron collisions.
A jet algorithm defines a map between the partonic phase space variables $\phi_n$ and $\phi_{n+1}$
and the jet momenta.
We write this map as
\bq
 \vec{j} \;\; = \;\; \vec{J}_{n,n}\left(\phi_n\right)
 & \mbox{and} &
 \vec{j} \;\; = \;\; \vec{J}_{n,n+1}\left(\phi_{n+1}\right).
\eq
The jet algorithm records in addition, which partons are clustered into a specific jet.
At LO, each parton is assigned to an individual jet, while at NLO we have on the $(n+1)$-particle phase space
the possibility that two partons are clustered into a single jet, while all remaining $(n-1)$ jets are formed
by one parton.

Eq.~(\ref{nlo_cross_section}) gives the perturbative NLO prediction for the total $n$-jet cross section.
It is clear that in a comparison between a theoretical 
perturbative calculation and experimental measurements the same jet algorithm has to be used.
The formulation of the NLO prediction in the form of eq.~(\ref{nlo_cross_section}) 
has the advantage, that once this formula is implemented into a numerical NLO program
the actual definition of the jet algorithm (or more general the infrared-safe observable) can 
easily be changed.

We may also consider differential cross sections.
The most detailed information is provided by the differential cross section which is differential
in all the independent jet momenta variables $\vec{j}$. We have with $t=\dim \vec{j}$
\bq
\label{fully_differential_cross_section}
 \frac{d^t\sigma_n^{\mathrm{NLO}}}{dj_1 ... dj_t}
 & = &
  \int d\phi_n \Theta_n\left(\phi_n\right)
 \left[ B\left(\phi_n\right) + V\left(\phi_n\right) + C\left(\phi_n\right) + I\left(\phi_n\right) \right] 
 \delta\left(\vec{j}-\vec{J}_{n,n}\left(\phi_n\right) \right)
 \nonumber \\
 & &
 +
 \int d\phi_{n+1}
  \Theta_n\left(\phi_{n+1}\right) \theta\left(\tau_n-\tau_n^{\mathrm{min}}\right) R\left(\phi_{n+1}\right) \delta\left(\vec{j} - \vec{J}_{n,n+1}\left(\phi_{n+1}\right) \right).
\eq
The fully differential cross section in eq.~(\ref{fully_differential_cross_section}) will be for reasonable input
parameters a non-negative function.
Negative values from eq.~(\ref{fully_differential_cross_section}) signal a breakdown of perturbation theory, for example
caused by a bad choice of renormalisation/factorisation scales or too extreme resolution cuts.

The delta distributions in the first line of eq.~(\ref{fully_differential_cross_section}) localise the integrand 
of the Born contribution to a point.
Thus for this contribution there are no integrals to be done.
Up to an additional convolution involving Altarelli-Parisi splitting functions hidden in the terms
$C$ and $I$ this is the case for all terms appearing in the first line of eq.~(\ref{fully_differential_cross_section}).
In the second line of eq.~(\ref{fully_differential_cross_section}) 
we have $(3n-2)$ delta distributions for hadron-hadron collisions and $(3n-4)$ delta distributions for 
electron-positron collisions.
The number of integrations is $(3n-1)$ for electron-positron collisions and
$(3n-1)+2$ for hadron-hadron collisions, where the extra two integrations refer to the integration
over the momentum fractions $x_a$ and $x_b$ in eq.~(\ref{leading_order_cross_section}).
In both cases we can use the delta distributions to eliminate all but three integrations.
Thus the real emission part involves 
an integration over a three-dimensional manifold.
The technical challenge is the efficient integration over this three-dimensional manifold.
We discuss techniques to do that in the next section.

\section{The matrix element method at NLO}
\label{sect:MEM_NLO}

In this section we discuss the new ingredients which appear when extending the matrix element method to NLO.
We will assume that we know how to handle the matrix element method at leading order.
In particular we assume that in the case of electron-positron collisions
there is a bijection between a jet momentum configuration $\vec{j}$
and a Born parton configuration $\phi_n$.
In the case of hadron-hadron collisions we assume that there is a bijection between
a jet momentum configuration $\vec{j}$ and the set $\{\phi_n,x_a,x_b\}$.

We make only very mild assumptions on the jet algorithm. We assume an infrared-safe jet algorithm, which provides
a jet function $\Theta_n$, jet momenta $\vec{J}_{n,m}$ and information which particles are clustered into which jets.
The jet algorithm may be provided as a computer code.
In particular we do not require that the jet algorithm is given by analytic formulae.

Our default transfer function is simply a product of delta distributions:
\bq
\label{default_transfer_function}
 W\left(\vec{x},\vec{y}\right)
 & = &
 \prod\limits_{i=1}^{r}
  \delta\left(x_i-y_i\right).
\eq
This is no real restriction:
Once we know how to solve the problem for the delta distributions, it is an easy generalisation to allow
Gaussian distributions for some or all variables.

\subsection{The strict fixed-order variant}
\label{strict_fixed_order}

We first discuss a variant based on strict fixed-order perturbation theory.
We take the experimentally measured variables $\vec{x}$ to be the (experimentally measured)
jet momenta configuration $\vec{j}^{\;\mathrm{exp}}$:
\bq
 \vec{x} & = & \vec{j}^{\;\mathrm{exp}}
\eq
We take the variables $\vec{y}$ to be the jet momenta configuration $\vec{j}^{\;\mathrm{theo}}$ from the perturbative NLO calculation.
The likelihood function at NLO is given by
\bq
\label{MEM_NLO}
 {\mathcal L}^{\mathrm{strict}}\left(\vec{x}|\vec{\alpha}\right)
 & = &
 \frac{1}{\sigma_n^{\mathrm{NLO}}}
 \left\{
 \int d\phi_n \Theta_n\left(\phi_n\right)
 \left[ B\left(\phi_n\right) + V\left(\phi_n\right) + C\left(\phi_n\right) + I\left(\phi_n\right) \right] 
 \delta\left(\vec{x} - \vec{J}_{n,n}\left(\phi_n\right) \right)
 \right.
 \nonumber \\
 & &
 \left.
 +
 \int d\phi_{n+1}
  \Theta_n\left(\phi_{n+1}\right) \theta\left(\tau_n-\tau_n^{\mathrm{min}}\right) R\left(\phi_{n+1}\right) \delta\left(\vec{x} - \vec{J}_{n,n+1}\left(\phi_{n+1}\right) \right)
 \right\}.
\eq
Eq.~(\ref{MEM_NLO}) follows directly from eq.~(\ref{fully_differential_cross_section})
and is nothing else than the NLO prediction to observe $n$ jets with jet momenta $\vec{x}$.
The technical challenge is a method to evaluate numerically the likelihood given in eq.~(\ref{MEM_NLO}).
To this aim we split eq.~(\ref{MEM_NLO}) into two parts:
\bq
 {\mathcal L}^{\mathrm{strict}}\left(\vec{x}|\vec{\alpha}\right)
 & = &
 {\mathcal L}^{\mathrm{strict}}_n\left(\vec{x}|\vec{\alpha}\right)
 +
 {\mathcal L}^{\mathrm{strict}}_{n+1}\left(\vec{x}|\vec{\alpha}\right),
\eq
with
\bq
 {\mathcal L}^{\mathrm{strict}}_n\left(\vec{x}|\vec{\alpha}\right)
 & = &
 \frac{1}{\sigma_n^{\mathrm{NLO}}}
 \int d\phi_n \Theta_n\left(\phi_n\right)
 \left[ B\left(\phi_n\right) + V\left(\phi_n\right) + C\left(\phi_n\right) + I\left(\phi_n\right) \right] 
 \delta\left(\vec{x} - \vec{J}_{n,n}\left(\phi_n\right) \right),
 \nonumber \\
 {\mathcal L}^{\mathrm{strict}}_{n+1}\left(\vec{x}|\vec{\alpha}\right)
 & = &
 \frac{1}{\sigma_n^{\mathrm{NLO}}}
 \int d\phi_{n+1}
  \Theta_n\left(\phi_{n+1}\right) \theta\left(\tau_n-\tau_n^{\mathrm{min}}\right) R\left(\phi_{n+1}\right) 
  \delta\left(\vec{x} - \vec{J}_{n,n+1}\left(\phi_{n+1}\right) \right).
\eq
The numerical evaluation of ${\mathcal L}^{\mathrm{strict}}_n(\vec{x}|\vec{\alpha})$ is basically a leading-order problem:
Given $\vec{x}$ we find the corresponding variables $\vec{z}=\{\phi_n\}$ (for electron-positron collisions)
or $\vec{z}=\{\phi_n,x_a,x_b\}$ (for hadron-hadron collisions), and evaluate the integrand
at this point including the appropriate Jacobian
\bq
 J_n^{\mathrm{B}}
 & = &
 \left| \det \left( \frac{\partial \vec{J}_{n,n}\left(\phi_n\right)}{\partial \vec{z}} \right) \right|^{-1}.
\eq
In most applications the Jacobian $J_n^{\mathrm{B}}$ is trivial.
Thus we obtain
\bq
\label{likelihood_n}
 {\mathcal L}^{\mathrm{strict}}_n\left(\vec{x}|\vec{\alpha}\right)
 & = &
 \frac{1}{\sigma_n^{\mathrm{NLO}}}
 \Theta_n\left(\phi_n\right)
 \left[ B\left(\phi_n\right) + V\left(\phi_n\right) + C\left(\phi_n\right) + I\left(\phi_n\right) \right] 
 J_n^{\mathrm{B}}.
\eq

The numerical evaluation of ${\mathcal L}^{\mathrm{strict}}_{n+1}(\vec{x}|\vec{\alpha})$ is more challenging. 
We explain in detail the case of electron-positron collisions.
The extension towards hadron-hadron collisions is straightforward, however the notation is more cumbersome, 
as we have to take in addition
the two integrations over the momentum fractions $x_a$ and $x_b$ into account.

\begin{figure}
\begin{center}
\includegraphics[scale=0.8]{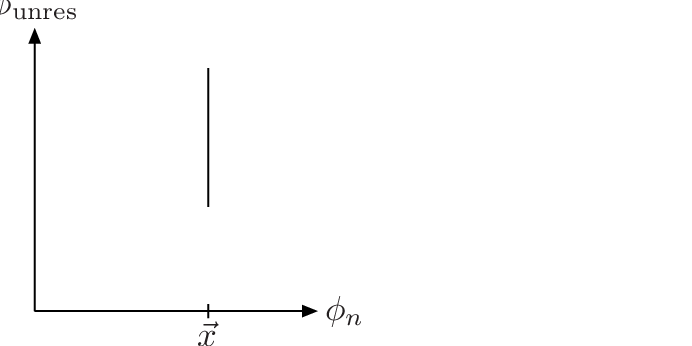}
\includegraphics[scale=0.8]{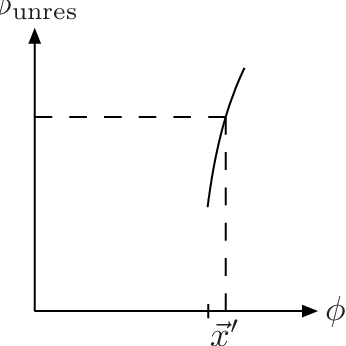}
\caption{\label{fig1} 
A sketch of of the region, which maps under the jet algorithm to the same jet configuration.
In the left figure we show the situation for the dedicated jet algorithm of \cite{Martini:2015fsa}.
The region is in the fibre above $\vec{x}$.
On the right figure we show the situation for an arbitrary jet algorithm, using the same coordinate system as before.
The region is now no longer within a single fibre. We may however still parametrise the three-dimensional manifold by
$\phi_{\mathrm{unres}}$.
}
\end{center}
\end{figure}

We may divide the $(n+1)$ particle phase space $\Phi_{n+1}$ into regions, where particles $i$ and $j$ are clustered into a jet
plus an irrelevant remainder region, which corresponds to the $(n+1)$-jet region or to regions with less than $n$ jets.
This defines a function $\theta_{i,j}$ which equals one if particles $i$ and $j$ are clustered into one jet and zero otherwise.
Thus, we may rewrite ${\mathcal L}^{\mathrm{strict}}_{n+1}(\vec{x}|\vec{\alpha})$ as
\bq
 {\mathcal L}^{\mathrm{strict}}_{n+1}\left(\vec{x}|\vec{\alpha}\right)
 & = &
 \frac{1}{\sigma_n^{\mathrm{NLO}}}
 \sum\limits_{(i,j)}
 \int d\phi_{n+1}
 \theta_{i,j}
  \Theta_n\left(\phi_{n+1}\right) \theta\left(\tau_n-\tau_n^{\mathrm{min}}\right) R\left(\phi_{n+1}\right) \delta\left(\vec{x} - \vec{J}_{n,n+1}\left(\phi_{n+1}\right) \right).
 \nonumber \\
\eq
In the region $\theta_{i,j}=1$ we use variables for $\Phi_{n+1}$, which correspond to a factorisation into $\Phi_n$ and 
$\Phi_{\mathrm{unres}}$.
In more mathematical terms we think about a fibre bundle, where the total space is given by $\Phi_{n+1}$, the base space by $\Phi_n$, the fibre by $\Phi_{\mathrm{unres}}$ and the projection by $\pi^{(\alpha)}$.
Given variables for the Born configuration $\phi_n$ and three variables for $\phi_{\mathrm{unres}}$ we may construct a point $\phi_{n+1}$
in $\Phi_{n+1}$.
We may think of these maps as the inverse of the Catani-Seymour projections $\pi^{(\alpha)}$ appearing in eq.~(\ref{projection_to_Born}). 
The appropriate formulae can be found in the literature \cite{Weinzierl:2001ny,Dinsdale:2007mf,Schumann:2007mg,Martini:2015fsa,Seth:2016hmv} and are not repeated here.
We denote this map as
\bq
 s & : & \Phi_n \times \Phi_{\mathrm{unres}} \rightarrow \Phi_{n+1},
 \nonumber \\
 & & \phi_{n+1} \; = \; s\left(\phi_n,\phi_{\mathrm{unres}} \right).
\eq
Let us first recapitulate the method proposed in \cite{Martini:2015fsa}.
There the authors define a dedicated jet algorithm based on the Catani-Seymour combination prescription
with the property that this jet algorithm combines the $(n+1)$-parton configuration $\phi_{n+1}$ exactly back to the original
Born configuration $\phi_n$.
All points of the region $\theta_{i,j}=1$ in $\Phi_{n+1}$,
which map under this jet algorithm to $\phi_n$ lie in the fibre above $\phi_n$.
This is illustrated in the left picture of fig.~(\ref{fig1}).
The integral over the radiation phase space is therefore a three-dimensional integral in the variables $\phi_{\mathrm{unres}}$.
The variables $\phi_n$ are fixed.

Let us now consider an arbitrary infrared-safe jet algorithm.
Keeping the same coordinate system with variables $\phi_n$ and $\phi_{\mathrm{unres}}$ as before
we have now the situation that the pre-image of the jet configuration $\vec{x}$ is no longer in the fibre above $\phi_n$.
One solution could be to find the analogue of the map $s$ for the specific jet algorithm under consideration.
However, this analytic inversion needs to be done for each jet algorithm separately 
and can be quite challenging or even impossible.
We are interested in a flexible method, which allows us to change the jet algorithm easily.
Thus, we look for a numerical solution.
The basic idea is to view the pre-image for an arbitrary jet algorithm as a deformation
of the pre-image of the jet algorithm advocated in \cite{Martini:2015fsa}.
Thus, we may still parametrise the three-dimensional manifold by $\phi_{\mathrm{unres}}$.
This is illustrated in the right picture of fig.~(\ref{fig1}).
In other words, there is a function
\bq
 f & : & \Phi_{\mathrm{unres}} \rightarrow \Phi_n
\eq
with
\bq
\label{condition_f}
 \vec{J}_{n,n+1}\left( s\left( f\left(\phi_{\mathrm{unres}}\right), \phi_{\mathrm{unres}} \right) \right)
 & = &
 \vec{x}.
\eq
The function value $f(\phi_{\mathrm{unres}})$ is given as the value $\phi_n = f(\phi_{\mathrm{unres}})$, which
satisfies eq.~(\ref{condition_f}). The point $\phi_n$ can be found by standard numerical methods,
for example Broyden's method \cite{Numerical_Recipes}.

With these prerequisites the likelihood ${\mathcal L}^{\mathrm{strict}}_{n+1}(\vec{x}|\vec{\alpha})$ becomes
\bq
\label{likelihood_n_plus_1}
 {\mathcal L}^{\mathrm{strict}}_{n+1}\left(\vec{x}|\vec{\alpha}\right)
 & = &
 \frac{1}{\sigma_n^{\mathrm{NLO}}}
 \sum\limits_{(i,j)}
 \int d\phi_{\mathrm{unres}}
 \theta_{i,j}
  \Theta_n\left(\phi_{n+1}\right) \theta\left(\tau_n-\tau_n^{\mathrm{min}}\right) R\left(\phi_{n+1}\right) 
  J^{\mathrm{R}},
\eq
where $\phi_{n+1}$ is given by
\bq
 \phi_{n+1}
 & = & 
 s\left(f\left(\phi_{\mathrm{unres}}\right),\phi_{\mathrm{unres}}\right).
\eq
The Jacobian $J^{\mathrm{R}}$ is given by
\bq
 J^{\mathrm{R}}
 & = &
 \left| \det \left( \frac{\partial \vec{J}_{n,n+1}\left(s\left(\phi_n,\phi_{\mathrm{unres}}\right)\right)}{\partial \vec{z}} \right) \right|^{-1}
\eq
and can be computed numerically by replacing derivatives with small finite differences.
For specific jet algorithms it might be possible to obtain an analytic formula for the Jacobian.

Let us summarise: The likelihood for an event within strict fixed-order perturbation theory is given by the sum
of the terms in eq.~(\ref{likelihood_n}) and eq.~(\ref{likelihood_n_plus_1}).
The evaluation of eq.~(\ref{likelihood_n}) is similar to an evaluation at leading-order.
On the other hand, the
evaluation of eq.~(\ref{likelihood_n_plus_1})
involves a three-dimensional integration.
For each integration point the method requires the numerical evaluation of the function $f(\phi_{\mathrm{unres}})$ 
and of the Jacobian $J^{\mathrm{R}}$.

The method described above is efficient to compute the likelihood 
${\mathcal L}_{n+1}(\vec{x}|\vec{\alpha})$ for an individual event $\vec{x}$ and 
model parameters $\vec{\alpha}$.
However, in practical applications we have to repeat this calculation a large number of times.
Indeed, from eq.~(\ref{likelihood_for_set_of_events}) we see that this calculation
has to be repeated $N$ times for a set of events $\{\vec{x}_1,...,\vec{x}_N\}$
and fixed model parameters $\vec{\alpha}$.
For large $N$ it might be more efficient to proceed as follows:
One performs for given model parameters $\vec{\alpha}$ a single NLO calculation
of the $n$-jet cross section $\sigma_n^{\mathrm{NLO}}$ with high Monte Carlo statistics
and bins the differential cross section in a multi-dimensional grid with bin sizes $\Delta x_i$.
The Monte Carlo statistics has to be high enough such that the statistical Monte Carlo error
in each bin is acceptable.
This NLO calculation can be performed either with the subtraction method according to
eq.~(\ref{subtracted_NLO}) or with the phase space slicing method according to eq.~(\ref{slicing_NLO}).
If one uses the subtraction method one encounters in particular for small bin sizes the problem, that
the subtraction terms end up in neighbouring bins, thus spoiling a cancellation within an individual bin.
Once this NLO calculation has been performed, we obtain the differential cross section for any
$\vec{x}$ by interpolation from the grid values.
The likelihood is then given by
\bq
 {\mathcal L}^{\mathrm{strict}}\left(\vec{x}|\vec{\alpha}\right)
 & = &
 \frac{1}{\sigma^{\mathrm{NLO}}}
 \frac{d^r\sigma_n^{\mathrm{NLO}}}{dx_1 ... dx_r}.
\eq

\subsection{The POWHEG-inspired variant}

Let us now discuss a small modification based on the POWHEG method \cite{Nason:2004rx,Frixione:2007vw,Alioli:2010xd}.
As before 
we take the experimentally measured variables $\vec{x}$ to be the experimentally measured
jet momenta configuration $\vec{j}^{\;\mathrm{exp}}$ and
we take the variables $\vec{y}$ to be the jet momenta configuration $\vec{j}^{\;\mathrm{theo}}$ from the 
theory calculation.
In the POWHEG-inspired variant we replace the strict fixed-order cross section $\sigma^{\mathrm{NLO}}$
with the POWHEG cross section $\sigma^{\mathrm{POWHEG}}$.

Let us first introduce the quantity $\bar{B}(\phi_n)$, given within the phase space slicing method by
\bq
\label{def_Bbar}
 \bar{B}\left(\phi_n\right) 
 & = &
 \left[ B\left(\phi_n\right) + V\left(\phi_n\right) + C\left(\phi_n\right) + I\left(\phi_n\right) \right] 
 +
 \int d\phi_{\mathrm{unres}} 
 \theta\left(\tau_n-\tau_n^{\mathrm{min}}\right)
 R\left(\phi_{n+1}\right).
\eq
Note that up to prefactors the quantity $\bar{B}(\phi_n)$ is very similar to 
the differential jet cross section in eq.~(\ref{fully_differential_cross_section})
and the NLO likelihood in eq.~(\ref{MEM_NLO}).
The essential difference is that in eq.~(\ref{def_Bbar}) the $(n+1)$-jet region is included.
The quantity $\bar{B}(\phi_n)$ can be computed with the methods discussed 
in section~\ref{strict_fixed_order}.

Let us further introduce a (shower-ordering) variable $p_\perp$, 
which equals in any singular limit the transverse momentum.
We define the Sudakov factor by
\bq
\label{def_Sudakov}
 \Delta\left(\phi_n,p_\bot^{\mathrm{min}}\right)
 & = &
 \exp\left(
 - \int d\phi_{\mathrm{unres}} \theta\left(p_\bot-p_\bot^{\mathrm{min}}\right) \frac{R\left(\phi_{n+1}\right)}{B\left(\phi_n\right)}
 \right).
\eq
Within the POWHEG-inspired approach we again write the likelihood as a sum of two terms
\bq
 {\mathcal L}^{\mathrm{POWHEG}}\left(\vec{x}|\vec{\alpha}\right)
 & = &
 {\mathcal L}^{\mathrm{POWHEG}}_n\left(\vec{x}|\vec{\alpha}\right)
 +
 {\mathcal L}^{\mathrm{POWHEG}}_{n+1}\left(\vec{x}|\vec{\alpha}\right),
\eq
where ${\mathcal L}^{\mathrm{POWHEG}}_n(\vec{x}|\vec{\alpha})$ is given by
\bq
\label{likelihood_n_powheg}
 {\mathcal L}^{\mathrm{POWHEG}}_n\left(\vec{x}|\vec{\alpha}\right)
 & = &
 \frac{1}{\sigma_n^{\mathrm{POWHEG}}}
 \Theta_n\left(\phi_n\right)
 \bar{B}\left(\phi_n\right)
 J_n^{\mathrm{B}}
 \Delta\left(\phi_n,p_\bot^{\mathrm{min}}\right).
\eq
and ${\mathcal L}^{\mathrm{POWHEG}}_{n+1}(\vec{x}|\vec{\alpha})$ is given by
\bq
\label{likelihood_n_plus_1_powheg}
 {\mathcal L}^{\mathrm{POWHEG}}_{n+1}\left(\vec{x}|\vec{\alpha}\right)
 & = &
 \frac{1}{\sigma_n^{\mathrm{POWHEG}}}
 \bar{B}\left(\phi_n\right)
 \sum\limits_{(i,j)}
 \int d\phi_{\mathrm{unres}}
 \theta_{i,j}
  \Theta_n\left(\phi_{n+1}\right) 
  \frac{R\left(\phi_{n+1}\right)}{B\left(\phi_n\right)} 
  J^{\mathrm{R}}
 \Delta\left(\phi_n,p_\bot\left(\phi_{n+1}\right)\right).
 \nonumber \\
\eq
Re-expanding the Sudakov factors one can show that the POWHEG cross section $\sigma^{\mathrm{POWHEG}}$
agrees with the strict fixed-order cross section $\sigma^{\mathrm{NLO}}$ up to NNLO terms.

\subsection{The variant based on identifying sub-jets}

In this sub-section we consider for concreteness the $k_\perp$-algorithm.
Up to now we always took 
the experimentally measured variables $\vec{x}$ to be the experimentally measured
jet momenta configuration $\vec{j}^{\;\mathrm{exp}}$.
A jet is made out of several particles and in this sub-section we investigate the possibility that
the transfer function is directly applied to the particles making up a jet.
Of course, our experimental and theoretical abilities are limited in this regard, but at 
next-to-leading accuracy the following approach seems reasonable:
We fix a (small) value $p_\bot^{\mathrm{min}}$ and
we divide the experimentally measured events into two sets:
The first set consists of all events, where no additional sub-jets are resolved
above $p_\bot^{\mathrm{min}}$.
The second set is the complement: at least one jet is resolved into sub-jets 
at a value $p_\bot>p_\bot^{\mathrm{min}}$.
Let us consider the second set in more detail.
We denote by $p_\bot^{\mathrm{split}}$ the largest value of $p_\bot$, where a jet is resolved
into two sub-jets.
We replace this jet by the two sub-jets.
In this way we arrive at an $(n+1)$ jet configuration.

The experimentally measured variables $\vec{x}$ are now given either by
an $n$-jet momenta configuration $\vec{j}_n^{\;\mathrm{exp}}$ (if the event belongs to the first set)
or by an $(n+1)$-jet momenta configuration $\vec{j}_{n+1}^{\;\mathrm{exp}}$
(if the event belongs to the second set).

If the event $\vec{x}$ corresponds to the first set (no sub-jets resolved above $p_\bot^{\mathrm{min}}$), we set
\bq
\label{likelihood_n_subjet}
 {\mathcal L}^{\mathrm{sub-jets}}_n\left(\vec{x}|\vec{\alpha}\right)
 & = &
 \frac{1}{\sigma_n^{\mathrm{POWHEG}}}
 \Theta_n\left(\phi_n\right)
 \bar{B}\left(\phi_n\right)
 J_n^{\mathrm{B}}
 \Delta\left(\phi_n,p_\bot^{\mathrm{min}}\right),
\eq
otherwise, if the event $\vec{x}$ belongs to the second set (i.e. there is a jet which is resolved into sub-jets above $p_\bot^{\mathrm{min}}$)
we set
\bq
\label{likelihood_n_plus_1_subjet}
 {\mathcal L}^{\mathrm{sub-jets}}_{n+1}\left(\vec{x}|\vec{\alpha}\right)
 & = &
 \frac{1}{\sigma_n^{\mathrm{POWHEG}}}
 \Theta_n\left(\phi_{n+1}\right) 
 \bar{B}\left(\phi_n\right)
 \frac{R\left(\phi_{n+1}\right)}{B\left(\phi_n\right)} 
 J^{\mathrm{B}}_{n+1}
 \Delta\left(\phi_n,p_\bot\left(\phi_{n+1}\right)\right).
\eq
The likelihood ${\mathcal L}^{\mathrm{sub-jets}}_n(\vec{x}|\vec{\alpha})$ 
in eq.~(\ref{likelihood_n_subjet}) can be interpreted as follows:
The quantity 
\bq
 \frac{\Theta_n(\phi_n) \bar{B}(\phi_n) J_n^{\mathrm{B}}}{\sigma_n^{\mathrm{POWHEG}}}
\eq
gives the NLO probability for the $n$-jet event $\vec{x}$.
This is multiplied by the Sudakov factor $\Delta(\phi_n,p_\bot^{\mathrm{min}})$, giving
the probability that no radiation above $p_\bot^{\mathrm{min}}$ occurs.
In the same way we may interpret eq.~(\ref{likelihood_n_plus_1_subjet}):
The quantity
\bq
  \frac{\Theta_n\left(\phi_{n+1}\right) R\left(\phi_{n+1}\right) J^{\mathrm{B}}_{n+1}}{\sigma_n^{\mathrm{POWHEG}}}
\eq
gives at the same order in perturbation theory the probability for a $(n+1)$-sub-jet event, which will
be clustered into a $n$-jet event.
This is multiplied by the Sudakov factor $\Delta(\phi_n,p_\bot(\phi_{n+1}))$,
giving the probability that no additional radiation with $p_\bot>p_\bot(\phi_{n+1})$ occurs.
The ratio $\bar{B}(\phi_n)/B(\phi_n)$ equals $1$ up to higher orders in perturbation theory.

In this paragraph we considered for concreteness 
the $k_\perp$-algorithm. For other jet algorithms one replaces in the definition of the Sudakov factor
in eq.~(\ref{def_Sudakov}) the argument of $\theta(p_\bot-p_\bot^{\mathrm{min}})$ with the appropriate resolution variable
from the jet algorithm.

The evaluations of $\bar{B}(\phi_n)$ and of the Sudakov factor $\Delta(\phi_n,p_\bot(\phi_{n+1}))$ 
require a three-di\-men\-sion\-al integration over the unresolved phase space.
However, it should be possible to extract at least the integrands of these integrals from the 
POWHEG BOX \cite{Alioli:2010xd}, 
making the POWHEG BOX a convenient starting point for an implementation of this variant.

\section{Conclusions}
\label{sect:conclusions}

In this paper we considered the extension of the matrix element method towards next-to-leading
order.
We discussed three variants.
Within the first two variants we take as the experimentally variables entering
the transfer function the experimentally measured jet momenta.
The first variant adheres to strict fixed-order perturbation theory and gives the likelihood
of an event exactly to next-to-leading order.
Within the second variant we replace the NLO cross section $\sigma^{\mathrm{NLO}}$
by the POWHEG cross section $\sigma^{\mathrm{POWHEG}}$.
Within the third variant we try to identify sub-jets above a scale $p_\bot^{\mathrm{min}}$.
If this is possible, the likelihood is given by the POWHEG-modified $(n+1)$-particle matrix element,
otherwise the likelihood is given by the POWHEG-modified $n$-particle matrix element.
 
\subsection*{Acknowledgements}

We would like to thank T. Martini and P. Uwer for useful comments on the manuscript.

\bibliography{/home/stefanw/notes/biblio}

\begin{thebibliography}{10}

\bibitem{Kondo:1988yd}
K.~Kondo,
\newblock J. Phys. Soc. Jap. {\bf 57}, 4126 (1988).

\bibitem{Kondo:1991dw}
K.~Kondo,
\newblock J. Phys. Soc. Jap. {\bf 60}, 836 (1991).

\bibitem{Kondo:1993in}
K.~Kondo, T.~Chikamatsu, and S.~H. Kim,
\newblock J. Phys. Soc. Jap. {\bf 62}, 1177 (1993).

\bibitem{Dalitz:1991wa}
R.~H. Dalitz and G.~R. Goldstein,
\newblock Phys. Rev. {\bf D45}, 1531 (1992).

\bibitem{Artoisenet:2010cn}
P.~Artoisenet, V.~Lemaitre, F.~Maltoni, and O.~Mattelaer,
\newblock JHEP {\bf 12}, 068 (2010), arXiv:1007.3300.

\bibitem{Brandt:2014tta}
O.~Brandt, G.~Gutierrez, M.~H. L.~S. Wang, and Z.~Ye,
\newblock Nucl. Instrum. Meth. {\bf A775}, 27 (2015), arXiv:1410.6319.

\bibitem{Fiedler:2010sg}
F.~Fiedler, A.~Grohsjean, P.~Haefner, and P.~Schieferdecker,
\newblock Nucl. Instrum. Meth. {\bf A624}, 203 (2010), arXiv:1003.1316.

\bibitem{Alwall:2010cq}
J.~Alwall, A.~Freitas, and O.~Mattelaer,
\newblock Phys. Rev. {\bf D83}, 074010 (2011), arXiv:1010.2263.

\bibitem{Campbell:2012cz}
J.~M. Campbell, W.~T. Giele, and C.~Williams,
\newblock JHEP {\bf 11}, 043 (2012), arXiv:1204.4424.

\bibitem{Campbell:2013hz}
J.~M. Campbell, R.~K. Ellis, W.~T. Giele, and C.~Williams,
\newblock Phys. Rev. {\bf D87}, 073005 (2013), arXiv:1301.7086.

\bibitem{Soper:2014rya}
D.~E. Soper and M.~Spannowsky,
\newblock Phys. Rev. {\bf D89}, 094005 (2014), arXiv:1402.1189.

\bibitem{Martini:2015fsa}
T.~Martini and P.~Uwer,
\newblock JHEP {\bf 09}, 083 (2015), arXiv:1506.08798.

\bibitem{Giele:1992vf}
W.~T. Giele and E.~W.~N. Glover,
\newblock Phys. Rev. {\bf D46}, 1980 (1992).

\bibitem{Giele:1993dj}
W.~T. Giele, E.~W.~N. Glover, and D.~A. Kosower,
\newblock Nucl. Phys. {\bf B403}, 633 (1993), hep-ph/9302225.

\bibitem{Keller:1998tf}
S.~Keller and E.~Laenen,
\newblock Phys. Rev. {\bf D59}, 114004 (1999), hep-ph/9812415.

\bibitem{Harris:2001sx}
B.~W. Harris and J.~F. Owens,
\newblock Phys. Rev. {\bf D65}, 094032 (2002), hep-ph/0102128.

\bibitem{Gao:2012ja}
J.~Gao, C.~S. Li, and H.~X. Zhu,
\newblock Phys. Rev. Lett. {\bf 110}, 042001 (2013), arXiv:1210.2808.

\bibitem{Boughezal:2015dva}
R.~Boughezal, C.~Focke, X.~Liu, and F.~Petriello,
\newblock Phys. Rev. Lett. {\bf 115}, 062002 (2015), arXiv:1504.02131.

\bibitem{Gaunt:2015pea}
J.~Gaunt, M.~Stahlhofen, F.~J. Tackmann, and J.~R. Walsh,
\newblock JHEP {\bf 09}, 058 (2015), arXiv:1505.04794.

\bibitem{Kunszt:1994mc}
Z.~Kunszt, A.~Signer, and Z.~Trocsanyi,
\newblock Nucl. Phys. {\bf B420}, 550 (1994), hep-ph/9401294.

\bibitem{Frixione:1996ms}
S.~Frixione, Z.~Kunszt, and A.~Signer,
\newblock Nucl. Phys. {\bf B467}, 399 (1996), hep-ph/9512328.

\bibitem{Catani:1997vz}
S.~Catani and M.~H. Seymour,
\newblock Nucl. Phys. {\bf B485}, 291 (1997), hep-ph/9605323.

\bibitem{Dittmaier:1999mb}
S.~Dittmaier,
\newblock Nucl. Phys. {\bf B565}, 69 (2000), hep-ph/9904440.

\bibitem{Phaf:2001gc}
L.~Phaf and S.~Weinzierl,
\newblock JHEP {\bf 04}, 006 (2001), hep-ph/0102207.

\bibitem{Catani:2002hc}
S.~Catani, S.~Dittmaier, M.~H. Seymour, and Z.~Trocsanyi,
\newblock Nucl. Phys. {\bf B627}, 189 (2002), hep-ph/0201036.

\bibitem{Weinzierl:2005dd}
S.~Weinzierl,
\newblock Eur. Phys. J. {\bf C45}, 745 (2006), hep-ph/0510157.

\bibitem{Frederix:2009yq}
R.~Frederix, S.~Frixione, F.~Maltoni, and T.~Stelzer,
\newblock JHEP {\bf 0910}, 003 (2009), arXiv:0908.4272.

\bibitem{Frixione:2011kh}
S.~Frixione,
\newblock JHEP {\bf 1109}, 091 (2011), arXiv:1106.0155.

\bibitem{Kosower:1998zr}
D.~A. Kosower,
\newblock Phys. Rev. {\bf D57}, 5410 (1998), hep-ph/9710213.

\bibitem{Campbell:1998nn}
J.~M. Campbell, M.~A. Cullen, and E.~W.~N. Glover,
\newblock Eur. Phys. J. {\bf C9}, 245 (1999), hep-ph/9809429.

\bibitem{Gehrmann-DeRidder:2005cm}
A.~Gehrmann-De~Ridder, T.~Gehrmann, and E.~W.~N. Glover,
\newblock JHEP {\bf 09}, 056 (2005), hep-ph/0505111.

\bibitem{Daleo:2006xa}
A.~Daleo, T.~Gehrmann, and D.~Maitre,
\newblock JHEP {\bf 04}, 016 (2007), hep-ph/0612257.

\bibitem{Somogyi:2006cz}
G.~Somogyi and Z.~Trocsanyi,
\newblock (2006), arXiv:hep-ph/0609041.

\bibitem{Nagy:2007ty}
Z.~Nagy and D.~E. Soper,
\newblock JHEP {\bf 09}, 114 (2007), arXiv:0706.0017.

\bibitem{Chung:2010fx}
C.~Chung, M.~Kramer, and T.~Robens,
\newblock JHEP {\bf 1106}, 144 (2011), arXiv:1012.4948.

\bibitem{Dittmaier:2008md}
S.~Dittmaier, A.~Kabelschacht, and T.~Kasprzik,
\newblock Nucl. Phys. {\bf B800}, 146 (2008), arXiv:0802.1405.

\bibitem{Czakon:2009ss}
M.~Czakon, C.~G. Papadopoulos, and M.~Worek,
\newblock JHEP {\bf 08}, 085 (2009), arXiv:0905.0883.

\bibitem{Gotz:2012zz}
D.~G{\"o}tz, C.~Schwan, and S.~Weinzierl,
\newblock Phys.Rev. {\bf D85}, 116011 (2012).

\bibitem{Bevilacqua:2013iha}
G.~Bevilacqua, M.~Czakon, M.~Kubocz, and M.~Worek,
\newblock JHEP {\bf 10}, 204 (2013), arXiv:1308.5605.

\bibitem{Stewart:2010tn}
I.~W. Stewart, F.~J. Tackmann, and W.~J. Waalewijn,
\newblock Phys. Rev. Lett. {\bf 105}, 092002 (2010), arXiv:1004.2489.

\bibitem{Stirling:1991ds}
W.~J. Stirling,
\newblock J. Phys. {\bf G17}, 1567 (1991).

\bibitem{Catani:1993hr}
S.~Catani, Y.~L. Dokshitzer, M.~H. Seymour, and B.~R. Webber,
\newblock Nucl. Phys. {\bf B406}, 187 (1993).

\bibitem{Ellis:1993tq}
S.~D. Ellis and D.~E. Soper,
\newblock Phys. Rev. {\bf D48}, 3160 (1993), arXiv:hep-ph/9305266.

\bibitem{Salam:2007xv}
G.~P. Salam and G.~Soyez,
\newblock JHEP {\bf 05}, 086 (2007), arXiv:0704.0292.

\bibitem{Cacciari:2008gp}
M.~Cacciari, G.~P. Salam, and G.~Soyez,
\newblock JHEP {\bf 04}, 063 (2008), arXiv:0802.1189.

\bibitem{Weinzierl:2001ny}
S.~Weinzierl,
\newblock JHEP {\bf 08}, 028 (2001), hep-ph/0106146.

\bibitem{Dinsdale:2007mf}
M.~Dinsdale, M.~Ternick, and S.~Weinzierl,
\newblock Phys. Rev. {\bf D76}, 094003 (2007), arXiv:0709.1026.

\bibitem{Schumann:2007mg}
S.~Schumann and F.~Krauss,
\newblock JHEP {\bf 03}, 038 (2008), arXiv:0709.1027.

\bibitem{Seth:2016hmv}
S.~Seth and S.~Weinzierl,
\newblock Phys. Rev. {\bf D93}, 114031 (2016), arXiv:1605.06646.

\bibitem{Numerical_Recipes}
W.~H. Press, S.~A. Teukolsky, W.~T. Vetterling, and B.~P. Flannery,
\newblock {\em Numerical Recipes in C} (Cambridge University Press, 1992).

\bibitem{Nason:2004rx}
P.~Nason,
\newblock JHEP {\bf 11}, 040 (2004), hep-ph/0409146.

\bibitem{Frixione:2007vw}
S.~Frixione, P.~Nason, and C.~Oleari,
\newblock JHEP {\bf 11}, 070 (2007), arXiv:0709.2092.

\bibitem{Alioli:2010xd}
S.~Alioli, P.~Nason, C.~Oleari, and E.~Re,
\newblock JHEP {\bf 06}, 043 (2010), arXiv:1002.2581.

\end{thebibliography}
\bibliographystyle{/home/stefanw/latex-style/h-physrev5}

\end{document}